2010 Fifth IEEE International Symposium on Electronic Design, Test & Applications# An ECG-on-Chip for Wearable Cardiac Monitoring Devices

C.J. Deepu, X.Y. Xu, X.D. Zou, L.B. Yao, and Y. Lian
Department of Electrical and Computer Engineering
National University of Singapore, Singapore
eleliany@nus.edu.sg**Abstract**

This paper describes a highly integrated, low power chip solution for ECG signal processing in wearable devices. The chip contains an instrumentation amplifier with programmable gain, a band-pass filter, a 12-bit SAR ADC, a novel QRS detector, 8K on-chip SRAM, and relevant control circuitry and CPU interfaces. The analog front end circuits accurately senses and digitizes the raw ECG signal, which is then filtered to extract the QRS. The sampling frequency used is 256 Hz. ECG samples are buffered locally on an asynchronous FIFO and is read out using a faster clock, as and when it is required by the host CPU via an SPI interface. The chip was designed and implemented in 0.35μm standard CMOS process. The analog core operates at 1V while the digital circuits and SRAM operate at 3.3V. The chip total core area is 5.74 mm$^2$ and consumes 9.6μW. Small size and low power consumption make this design suitable for usage in wearable heart monitoring devices.

**Keywords**: Electrocardiography, QRS detection, ECG-on-Chip, Low Power design, Wearable devices.## 1 Introduction

Remote monitoring of ECG and other vital physiological signals continuously is becoming increasingly important as it can significantly reduce the costs and risks involved in personal healthcare. The main challenge involved in the development of a constant remote health monitoring system, as shown in Figure 1, is the development of low power and compact wearable sensors which can acquire, process and wirelessly transmit these signals to a monitoring device. A high level of integration is required to minimize the size and cost of such a sensor. Since the wireless transceiver is the major source of power consumption in such a system, it is desirable to do most of the signal processing tasks like ECG filtering and QRS detection locally, as this will reduce number of bits to be transmitted. There are several attempts towards the design of various discrete and integrated components of such systems [1-5]. In [1-2] an integrated sensor interface chip which does signal acquisition and data conversion for this application is presented.

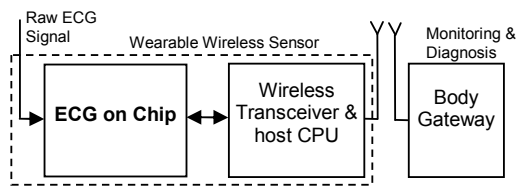

Fig.1 Typical wearable wireless heart monitoring system

In this paper, we present a fully integrated, low power chip for ECG signal processing in wearable sensors. The chip combines an instrumentation amplifier with programmable gain stage, a 12-bit successive approximation (SAR) ADC, a novel mathematical morphology [6] based QRS detector, a central control unit with 8K on-chip SRAM and 2 SPI interfaces. The chip was designed and implemented in 0.35μm standard CMOS process. The chip total core area is 5.74 mm$^2$ and consumes 9.6μW. Small size and low power consumption make this chip suitable for usage in wearable heart monitoring devices.

This paper is organized as follows. In Section II, the overall architecture of the proposed ECG chip is described. In Section III, the details of hardware implementation of individual blocks are given. Simulation results and power consumption details are presented in Section IV. Concluding remarks are given in Section V.

## 2 System Architecture

The system architecture of proposed ECG-on-Chip is shown in Figure 2. An instrumentation amplifier with programmable gain stage and embedded band-pass filtering function is used for signal acquisition. The amplified ECG signal is digitized by a successive approximation ADC. The sampling frequency of ADC used is 256 Hz. The QRS detection block extracts the QRS complexes from the ECG Signal. Traditionally, ECG noises such as baseline wander, electromyographic interference (EMG), power line noise, and motion artifacts are removed by digital filters in order to extract QRS. However, such approach is not well suited for wearable devices due to large area and power consumption. This chip uses a novel, low complexity QRS detector based on multi-scale mathematical morphology [6] achieving very low power consumption. The QRS detector, as shown in Figure 3, computes the user's heart rate by keeping

978-0-7695-3978-2/10 $26.00 © 2010 IEEE
DOI 10.1109/DELTA.2010.43

225

track of the R-R interval and is updated once every 10s. This block has a dedicated SPI interface to readout the R-R interval.

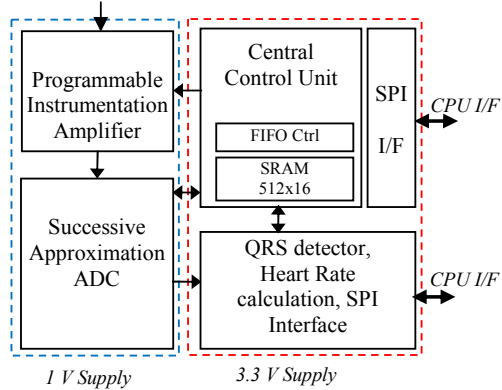

Fig.2 ECG-on-Chip Architecture

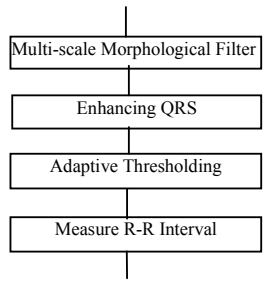

Fig.3 Multiscale-Morphology based QRS detection

The central control unit (CCU) decodes the CPU commands and generates relevant control signals for each individual block in the chip. It contains an asynchronous FIFO, which is used to interface the chip with host CPU having faster clocks. The ECG data from ADC and heart rate from QRS detector is continuously written into the asynchronous FIFO @ 256 Hz and is read out by the host CPU at a higher clock speed once the FIFO is full. This interfacing mechanism saves system power as CPU can switch to sleep mode when the data is being written into the FIFO.

As shown in Figure 2, the analog core operates at 1V supply while the digital circuits and SRAM operate at 3.3V. A lower supply voltage helps to save power in the analog part. The analog circuits are custom designed and digital circuits are designed using standard cell based design methodology. Level converters are implemented on chip, in order to interface between different voltage domains.

## 3 Hardware Implementation

### 3.1 Instrumentation Amplifier

The analog front-end is in charge of the noise suppression, signal conditioning and amplification. As shown in Figure 4, it consists of two stages. The first stage is a low noise front-end amplifier with band-pass function. The second stage is a programmable gain amplifier (PGA) adopting flip-over-capacitor technique [1-2]. Due to the ultra large resistance generated by the pseudo-resistors, it usually takes very long time for the front-end amplifier to settle down when power is just applied. In this design, two switches S1 and S2 are added. A reset signal will switch them on during startup of the system, which speeds up the settling of the front-end amplifiers significantly.

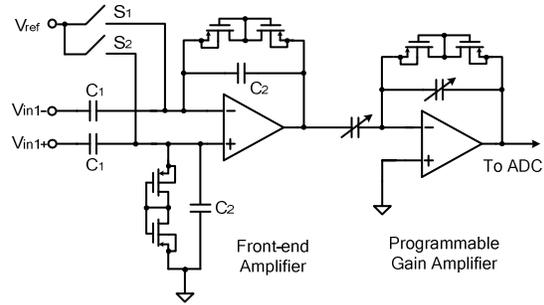

Fig.4. Schematic of the low noise front-end amplifiers

### 3.2 Successive Approximation ADC

SAR ADC is known for its moderate accuracy, moderate speed, and low power overhead, thus it is well suited for this application. Figure 5 depicts the architecture of the proposed SAR ADC [1-2]. In each conversion cycle, the analog input is sampled through a bootstrapped switch and held on the capacitive DAC. Driven directly by the preceding buffer stage without the need of additional hold amplifier, the employed open-loop S/H achieves fast settling and small offset error at low power cost. The captured analog signal is then compared with a fixed reference REF and level-shifted by the DAC in a sequence of binary search. The fixed REF largely eliminates the dynamic offset associated with the comparator. The use of an array of capacitors to implement the DAC assures excellent matching and noise performance. The SAR logic and timing sequence are driven by an on-chip crystal oscillator that features both low jitter and low power consumption. The obtained digital codes are level-converted and passed to the CCU and QRS detection modules for further processing.

### 3.3 QRS Detector

This block filters the incoming ECG signal to remove the noise components and to locate QRS complexes and estimate R-R interval.

The proposed morphological filter is a modified version of [6], which is used to remove noises in the ECG signal and detects QRS complex. The filter consists of a pair of opening and closing operations as shown in Figure 6. Opening and Closing operations behave as a filter that suppresses peaks and valleys and are implemented using dilation and erosion operators. The averaged output of opening and closing operations is subtracted from the original input to remove the wandering baseline drift.

The typical duration of QRS complex is around 0.06-0.1s which corresponds to a length of 25 samples, using a sampling rate of 256 Hz. Therefore the dilation and erosion operations were done over a duration of 25 samples. The input ECG samples are serially loaded into the shift register and is added (or subtracted, for erosion) with the corresponding structure element $g(x)$[6]. The results are then continuously compared using a comparator tree to find the minimum (or maximum, for erosion) for the dilation/erosion operation.

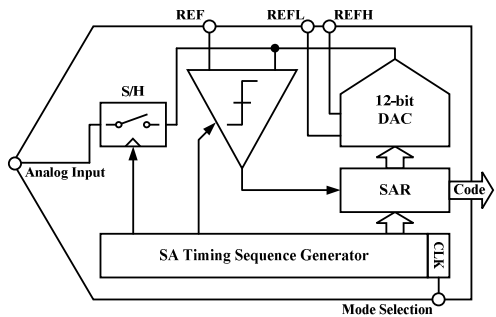

Fig 5. System architecture of the SAR ADC.

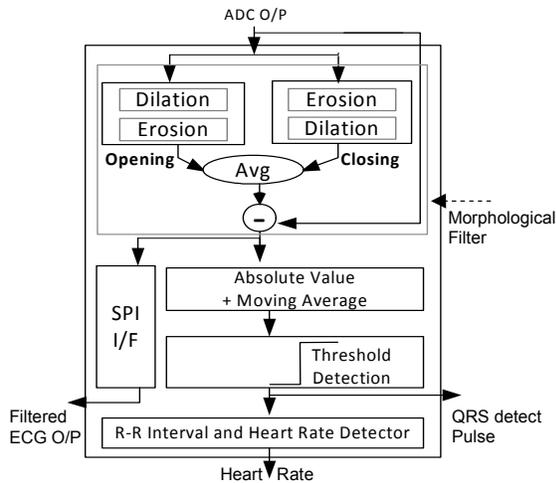

Fig 6. QRS detection block diagram

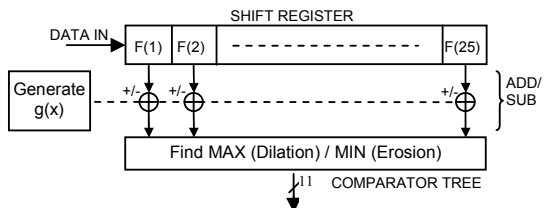

Fig 7. Dilation/Erosion block structure

The filtered signal is further smoothened to remove the impulse noise using a moving average filter. A simple serial structure is used to implement the moving average filter. This output is continuously monitored by a threshold detector in order to detect the R-peak. A new peak is detected by comparing the received signal against an adaptive threshold. Every time a new peak is detected, the current threshold is updated and it corresponds to the newly detected peak.

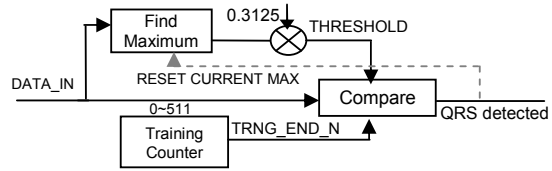

Fig 8 Adaptive threshold detector

The R-R interval of the received ECG signal is measured using a binary counter which counts the number of clocks between R peaks. Furthermore, the heart rate is calculated by counting the number R peaks in the last 60s. A parallel-to-serial converter is included for sending out the heart rate through SPI interface.

### 3.4   Central Control Unit(CCU)

CCU block, as shown in Figure 9, generates interface control signals for all blocks in the chip, based on the host CPU commands. It does data framing, CPU interrupt generation etc. based on state machine control signals. This block also includes an asynchronous FIFO with 8 Kb buffer, in-order to interface the chip with different type of host CPUs. Data from ADC and QRS block is continuously written into the FIFO at the sampling frequency. The FIFO write, read controllers generate status signals such as "full", "nearly full", "empty", "nearly empty" based on the FIFO status. Therefore, as the FIFO gets filled, CCU state machine, as shown in Figure 10, interrupts the host CPU and request a read operation. FIFO read and write pointers are implemented as binary counters. In order to generate the FIFO status signals by pointer comparison, the binary pointers are converted to gray code equivalent and synchronized across clock domains. This way the FIFO pointers are protected from accidental overflow, underflow conditions that might occur due to meta-stability while comparing pointers across clock domains.

### 3.5   SPI Interface

The proposed device communicates with external microcontroller via a duplex SPI slave interface, whose structure is shown in Figure 11. The data link carries the ECG and QRS codes, along with the internal FIFO status flags. The command link delivers the control vector from the external microcontroller to the internal control registers. In addition to the driving clocks for the I/O shift registers, the clock extraction circuit also extracts the CCU clocks that pace the FIFO reading thread.

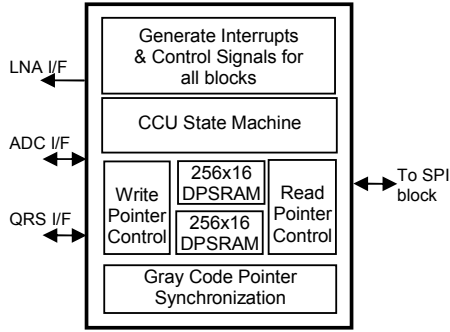

Fig 9 Central Control Unit (CCU)

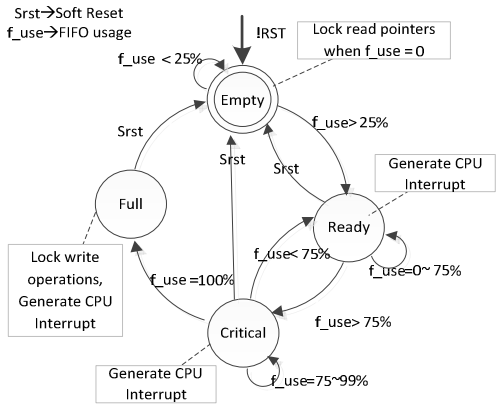

Fig 10 CCU State Machine

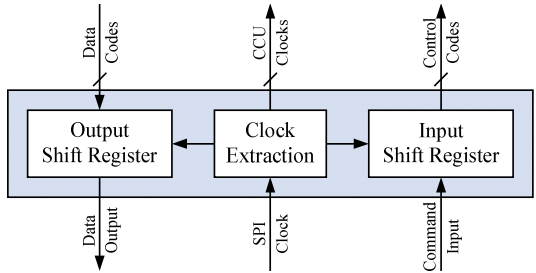

Fig 11. Block diagram of the duplex SPI slave interface.

### 3.6 Simulation results

The design was implemented by 0.35μm standard CMOS process. The total core area is 5.74 mm2 and layout is shown in Figure 12. The power consumption of the chip is 0.85μW@1V for analog circuits and ADC, 1.1 μW@3.3V for QRS detector, and 7.6 μW@3.3V for CCU and other digital circuits (assuming 1MHz host CPU, but excluding SRAM).

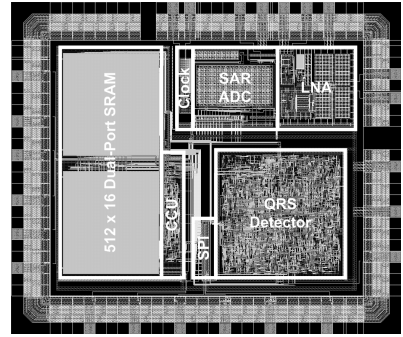

Fig.12. Layout of ECG-on-Chip

## 4 Conclusion

This paper has presented the design of an ECG-on-Chip for wearable ECG devices. Low power ECG acquisition and ADC circuits were implemented and tested. A novel, low complexity architecture for QRS detection, with high detection rate, has been proposed and implemented. A central chip control unit with on-chip SRAM buffer is developed for interfacing the chip with a host CPU. The design consumes 9.6μW and has a core area of 5.74mm$^2$ in 0.35μm standard CMOS process. High level of integration and low power consumption make this design suitable for development of low cost wearable cardiac monitoring devices.

## 5 Acknowledgements

This work was supported by the research grants 052-118-0057, 052-118-0060 and 072-118-0030 from Singapore Agency for Science Technology and Research (A*STAR).